\documentclass[aip,jap,a4,reprint,twoside,floatfix,superscriptaddress]{revtex4-1} 

\pdfoutput=1
\usepackage{graphicx}
\usepackage{amsmath}
\usepackage{amssymb}
\usepackage{amsfonts}
\usepackage{dcolumn}
\usepackage{epsfig}
\usepackage{subfigure}
\usepackage{bm}

\begin{document}

\title{\bf Room temperature multiferroicity in orthorhombic LuFeO$_3$ }

\author {Ujjal Chowdhury} \affiliation {Nanostructured Materials Division, CSIR-Central Glass and Ceramic Research Institute, Kolkata 700032, India}
\author{Sudipta Goswami} \affiliation {Nanostructured Materials Division, CSIR-Central Glass and Ceramic Research Institute, Kolkata 700032, India}
\author {Dipten Bhattacharya}
\email{dipten@cgcri.res.in} \affiliation {Nanostructured Materials Division, CSIR-Central Glass and Ceramic Research Institute, Kolkata 700032, India}
\author{Jiten Ghosh} \affiliation {Materials Characterization Division, CSIR-Central Glass and Ceramic Research Institute, Kolkata 700032, India} 
\author {Soumen Basu} \affiliation {Department of Physics, National Institute of Technology, Durgapur 713209, India}
\author {Samya Neogi} \affiliation {Department of Physics, National Institute of Technology, Durgapur 713209, India}

\date{\today}

\begin{abstract}
From the measurement of dielectric, ferroelectric, and magnetic properties we observe simultaneous ferroelectric and magnetic transitions around $\sim$600 K in orthorhombic LuFeO$_3$. We also observe suppression of the remanent polarization by $\sim$95\% under a magnetic field of $\sim$15 kOe at room temperature. The extent of suppression of the polarization under magnetic field increases monotonically with the field. These results show that even the orthorhombic LuFeO$_3$ is a room temperature multiferroic of type-II variety exhibiting quite a strong coupling between magnetization and polarization.
\end{abstract}

\maketitle

The magnetoelectric multiferroics with strong cross-coupling between ferroelectric and magnetic order parameters have attracted a lot of attention during the last one decade because of their potential in radically enhancing the functionalities of the spintronics-based devices for many, including bio-medical, applications.\cite{Wang} A room temperature multiferroic, for obvious reasons, is always the most sought-after compound. Apart from BiFeO$_3$, other systems such as CuO [Ref. 2] and KBiFe$_2$O$_5$ [Ref. 3] have also been identified as potential high temperature multiferroic compounds. In recent time, hexagonal LuFeO$_3$ is found to exhibit multiferroicity at room temperature.\cite{Xu} Coexistence of ferrimagnetic and ferroelectric orders was reported in orthorhombic LuFe$_{1-x}$Mn$_x$O$_3$ as well.\cite{Qin} However, in none of these recent work direct measurement of the multiferroic coupling has been attempted. In this Letter, we show that even the pure orthorhombic-LuFeO$_3$, in bulk form, exhibits large magnetoelectric multiferroic coupling at room temperature. The remanent polarization ($P_r$) is suppressed by $\sim$95\% under a magnetic field of $\sim$15 kOe. The extent of suppression of $P_r$ increases monotonically with the increase in magnetic field. The ferroelectric and magnetic transitions are simultaneous around $\sim$600 K indicating magnetic structure driven ferroelectricity. This observation signfies that the orthorhombic LuFeO$_3$ is a multiferroic of type-II variety.

The structurally nonpolar orthoferrites such as SmFeO$_3$, YFeO$_3$, LuFeO$_3$ etc. exhibit small yet finite ferroelectric polarization because of noncollinear spin structure with canted antiferromagnetic order.\cite{Lee,Shang,White} This magnetic order induces a spin current\cite{Katsura} via spin-orbit-coupling driven antisymmetric exchange interaction along the Fe$^{3+}$-O$^{2-}$-Fe$^{3+}$ pathway. The spin current ($\vec{j_s}$) in such a noncollinear magnetic structure breaks the centrosymmetry of the electronic charge density distribution and yields a finite polarization ($\vec{P}$ $\propto$ $\vec{j_s}$).\cite{Katsura,Mostovoy} These compounds, where magnetism drives ferroelectricity, belong to the type-II category of the multiferroics. Interestingly, unlike the well-known perovskite manganites such as TbMnO$_3$ or DyMnO$_3$, the rare-earth orthoferrites exhibit the ferroic transitions at well above the room temperature. In fact, for many of these orthoferrites, the magnetic and ferroelectric transition points (T$_N$ and T$_C$, respectively) are found to be varying within $\sim$600-700 K.\cite{White} However, the extent of ferroelectric polarization and, more importantly, the multiferroic coupling are yet to be determined from direct electrical measurements under a magnetic field or vice versa. We show here that the bulk orthorhombic LuFeO$_3$ exhibits a finite ferroelectric polarization as well as a strong multiferroic coupling testified by a sharp drop in the remanent polarization under a magnetic field at room temperature.

\begin{figure}[!ht]
  \begin{center}
    \includegraphics[scale=0.30]{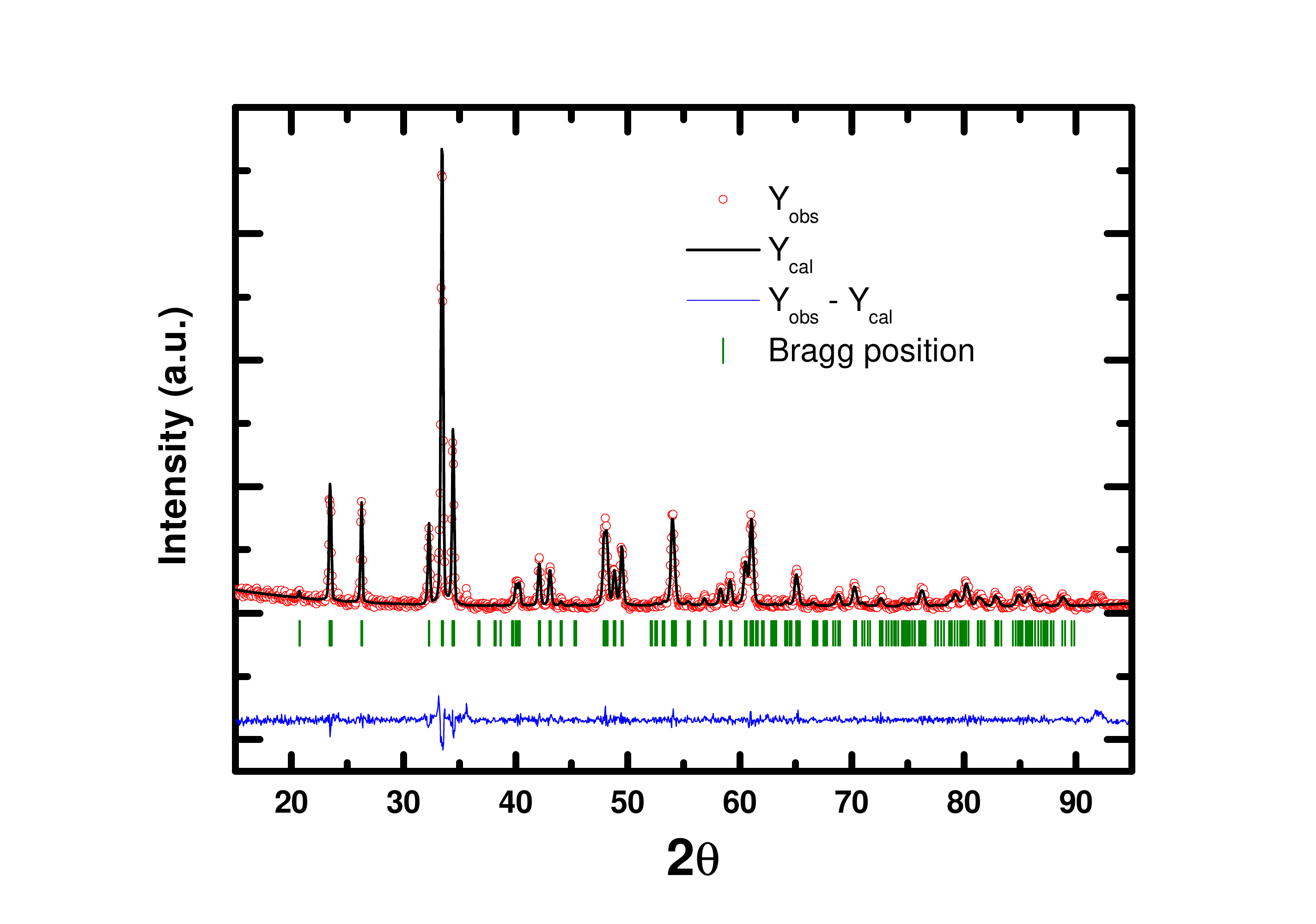} 
    \end{center}
  \caption{The powder x-ray diffraction pattern and its refinement by FullProf for bulk orthorhombic LuFeO$_3$. }
\end{figure}

The bulk LuFeO$_3$ pellets were synthesized by the standard solid-state reaction process using Lu$_2$O$_3$(99.99$\%$), Fe$_2$O$_3$ (99$\%$) as precursor. The raw powders were mixed in stoichiometric ratio for 24h with the help of a high energy ball mill with zirconia balls in deionized water. The mixtures were calcined at 1373 K in air for 4h. Then, the calcined product was again ball milled for 12h. An organic binder (PVA) was mixed with the sample by 5 wt$\%$ and pressed under uniaxial pressure of 80 MPa to prepare the pellets. Finally, the pellets were sintered at 1473 K for 10h. The sintered product was characterized by room temperature x-ray diffraction and scanning electron microscopy. A representative scanning electron microscopy image of the sample microstructure is shown in the supplementary document.\cite{supplementary} The microstructure appears to be quite dense. The x-ray diffraction pattern was refined by Fullprof to extract the structural noncentrosymmetry and other details. For measuring the dielectric and ferroelectric properties, we used a disk-shaped sample of diameter $\sim$8 mm and thickness $\sim$1 mm. A layer of silver paste was applied as the electrode on the top and bottom surfaces of the sample in two-probe configuration. The paste was cured at $\sim$500$^o$C for proper sample-electrode adhesion. The dielectric properties were measured by an LCR meter (HIOKI Technologies) across the frequency range 42 Hz to 2 MHz. The magnetic properties were measured by a Vibrating Sample Magnetometer (VSM)(LakeShore Cryotronics, Model 7407). The remanent hysteresis loops were measured under zero as well as different magnetic fields by Precision Material Analyzer (Radiant Technologies).

\begin{figure}[!ht]
  \begin{center}
    \includegraphics[scale=0.42]{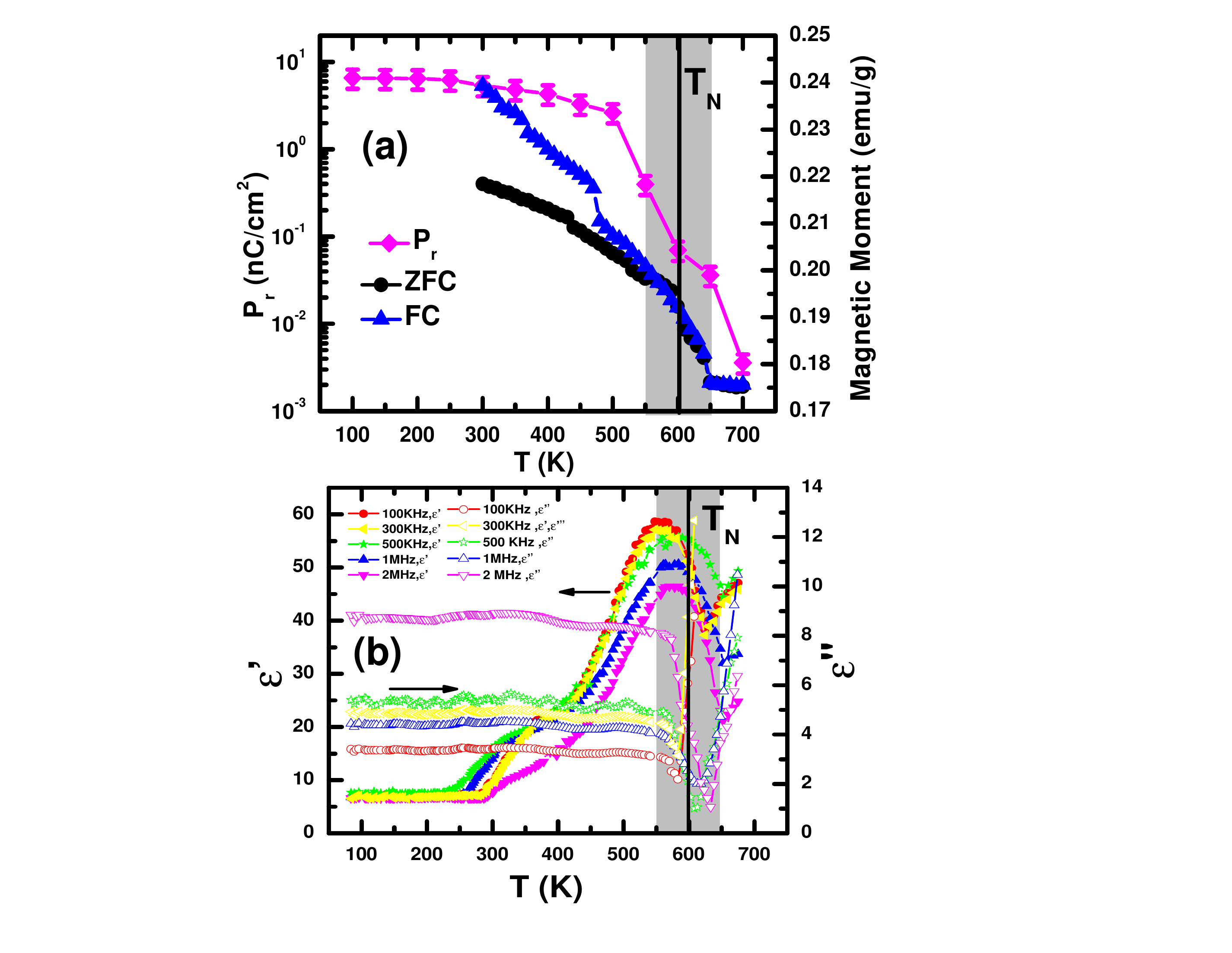} 
    \end{center}
  \caption{(color online) (a) The ZFC and FC magnetization versus temperature plots across 300-700 K showing the magnetic transition around $\sim$600 K (right y-axis); measurements were carried out under $\sim$100 Oe field; also shown are the remanent polarization versus temperature (left y-axis); (b) real and imaginary dielectric permittivity versus temperature patterns; anomaly around $\sim$600 K is clearly visible both in the real and imaginary parts. }
\end{figure}

In Fig. 1, we show the room temperature x-ray diffraction pattern and its refinement. The data were refined by considering the orthorhombic structure with space group $\textit{Pbnm}$. Attempt has also been made to refine the data by considering hexagonal \textit{P6$_3$cm} space group. A comparative analysis of the refinement is given in the supplementary document.\cite{supplementary} It appears that the structure is indeed orthorhombic. The structural details such as the lattice parameters, bond length, angle, distortion etc together with the fit statistics are given in the supplementary document.\cite{supplementary} The unit cell noncentrosymmetry, however, turns out to be zero as expected for $\textit{Pbnm}$ space group. The small ferroelectric polarization observed here then arises from the noncollinear magnetic structure via spin supercurrent induced between the spins coupled antisymmetrically as $\vec{\bf{S_1}} \times \vec{\bf{S_2}}$.\cite{Lee,Shang,Katsura,Mostovoy} In Fig. 2a, we show the zero-field-cooled (ZFC) and field-cooled (FC) magnetization versus temporature plot across 300-700 K. The magnetic transition temperature $T_N$ $\sim$600 K could be identified from the plot. The transition is, of course, a bit broader, possibly, because of the presence of structural/electronic inhomogeneities. The transition zone ($\sim$100 K) is marked by the shaded region. Also shown in Fig. 2a is the remanent polarization ($P_r$) measured across a wide temperature range 100-700 K. Fig. 2b shows the real and imaginary parts of the dielectric permittivity ($\epsilon'$, $\epsilon"$) for a few frequencies. In fact, in order to show the intrinsic dielectric response, we plot in Fig. 2b the $\epsilon'$ and $\epsilon"$ measured at higher frequency. The anomaly around the magnetic transition point ($\sim$600 K) is clearly visible signifying a coupling between dielectric and magnetic properties. The complex plane impedance spectra describing the dielectric relaxation across the entire frequency range of 42 Hz-2 MHz are shown in the supplementary document.\cite{supplementary} Separate relaxation regimes could be noticed for interface and bulk dielectric property only in the higher temperature range ($\ge$400 K). By fitting the impedance spectra with an appropriate equivalent circuit model the intrinsic bulk resistance ($R_b$), capacitance ($C_b$), and the relaxation exponent ($\beta$) were determined across 100-700 K.\cite{supplementary} The influence of the interface dielectric relaxation could be eliminated by the equivalent circuit model. The plot of $R_b$, $C_b$, and $\beta$ with temperature, shown in the supplementary document, also depicts clear anomaly around the $T_N$. Interestingly, $P_r$ too, shows the transition around $T_N$. It is important to mention here that though the transition is broad, the onset and completion of the transition for all the dielectric, ferroelectric, and magnetic properties take place around $\sim$600 K with clear overlap in the transition zone. This observation signifies that like other type-II multiferroic systems, orthorhombic LuFeO$_3$ is a type-II multiferroic exhibiting simultaneous ferroelectric and magnetic transitions. It supports the notion that the ferroelectric order here is originating from the canted noncollinear magnetic structure. 

\begin{figure}[!ht]
  \begin{center}
    \includegraphics[scale=0.25]{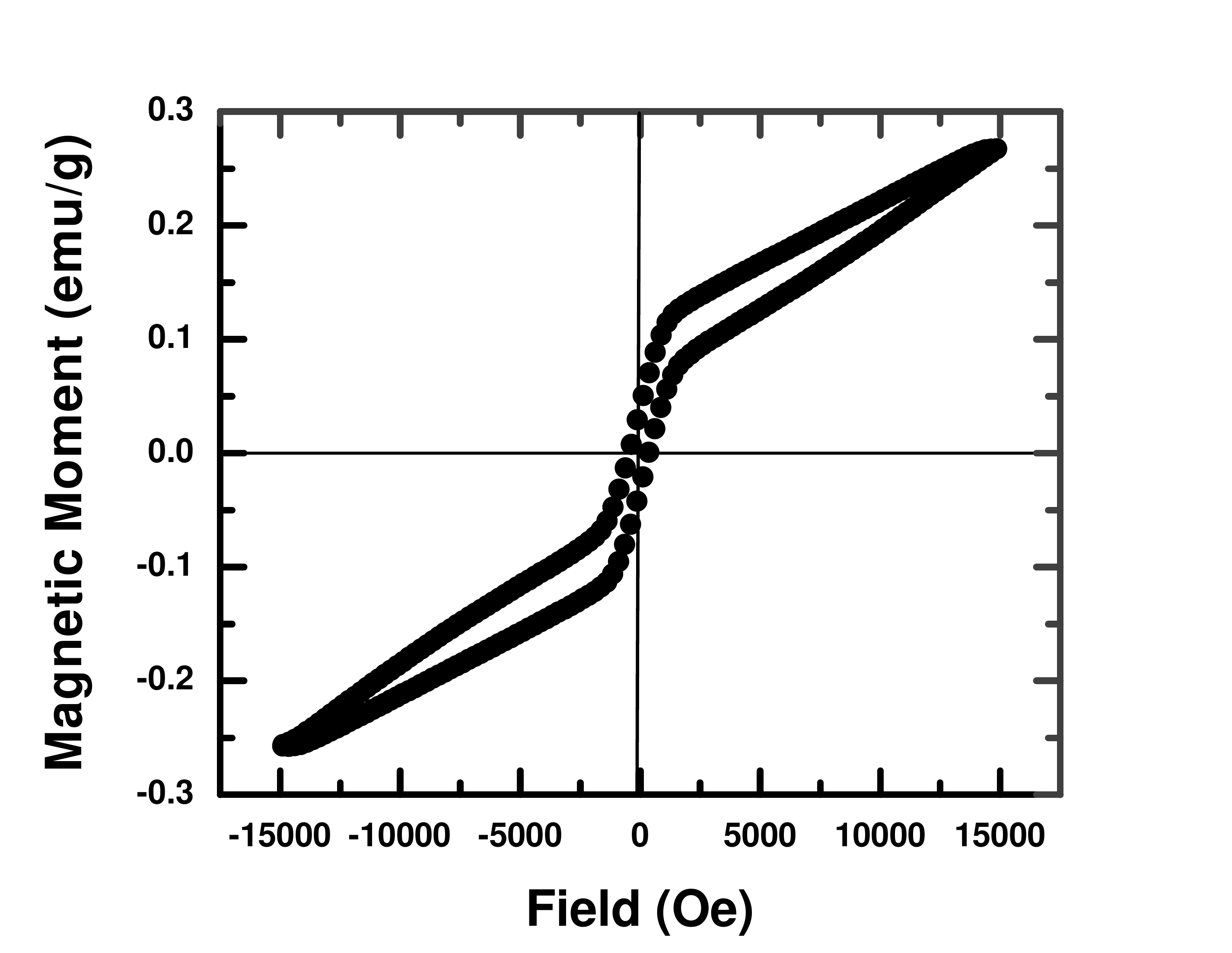} 
    \end{center}
  \caption{The magnetic hysteresis loop measured at room temperature. }
\end{figure}

\begin{figure}[!ht]
  \begin{center}
    \includegraphics[scale=0.25]{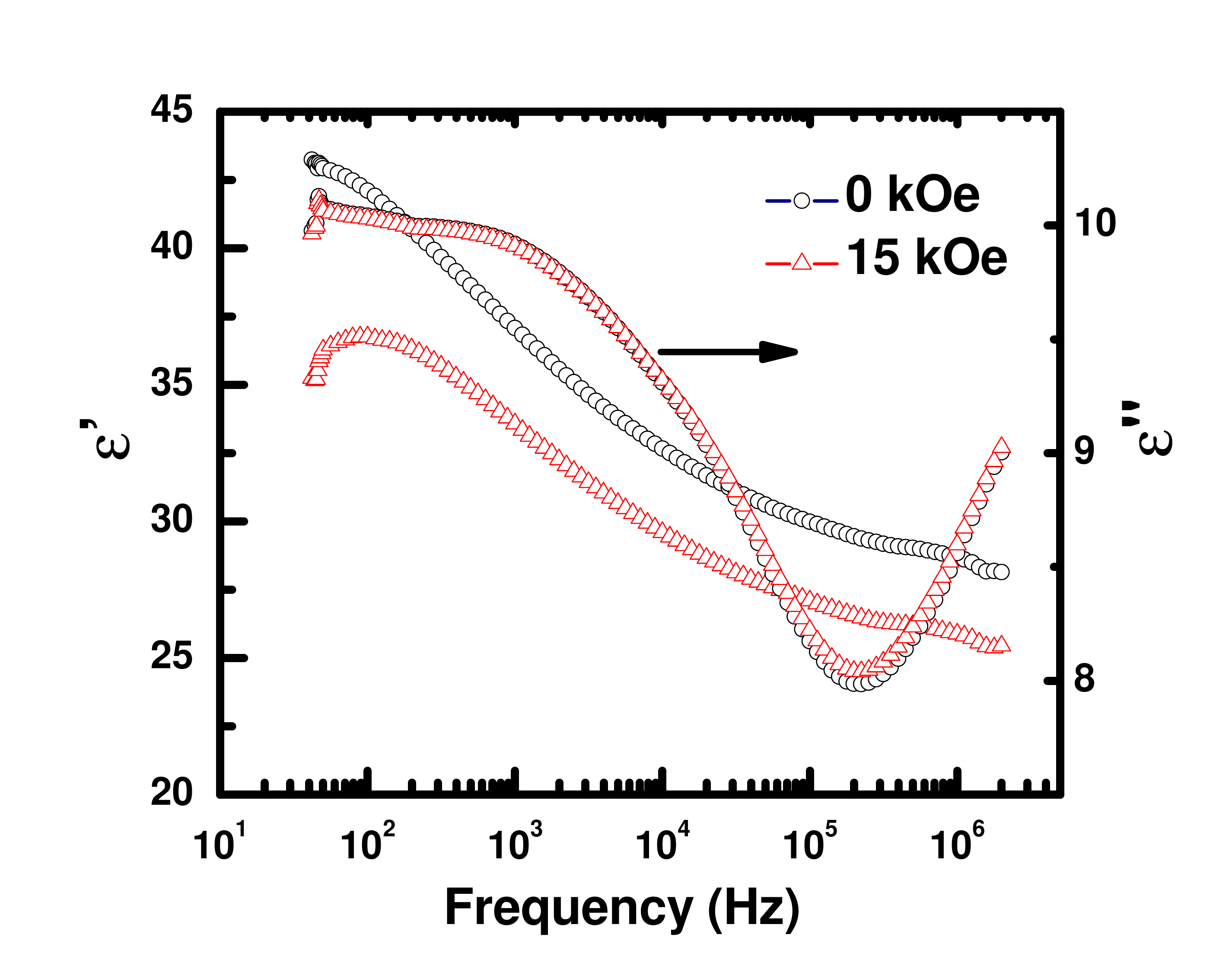} 
    \end{center}
  \caption{The real and imaginary parts of the dielectric permittivity as a function of frequency measured under zero and $\sim$15 kOe magnetic field at room temperature. The influence of magnetic field on the real part is far stronger than that on the imaginary part. }
\end{figure}

Fig. 3 shows the room temperature magnetic hysteresis loop. The sample exhibits a magnetization of $\sim$0.016 $\mu_B$/Fe at a magnetic field of $\sim$15 kOe at room temperature. Weak ferromagnetism is expected in canted noncollinear spin structure. The magnetization observed here is, of course, a bit smaller than what has been estimated\cite{Lee} in isostructural SmFeO$_3$. In Fig.4, we show the magnetodielectric effect observed at room temperature. Interestingly, the suppression of the real part of the dielectric permittivity under a magnetic field is quite substantial ($\sim$10\%) whereas the change in the imaginary part is rather small ($\le$0.6\%). It appears that the magnetodielectric effect essentially results from the intrinsic magnetocapacitive component. Similar observation has been made in the multiglass La$_2$NiMnO$_6$ system as well.\cite{Choudhury}   

In order to verify the multiferroicity in orthorhombic LuFeO$_3$ even further, we also measured the influence of magnetic field on intrinsic remanent polarization. The polarization versus electric field loop measured by a simple triangular voltage pulse - normally used for a conventional ferroelectric system with large polarization - yields a loop characteristic of lossy dielectric systems. The ordinary loop suffers from the contribution from resistive leakage as well as nonswitchable polarization which gives a wrong estimate of the polarization. The systems exhibiting small polarization need a different type of protocol for determination of the intrinsic switchable ferroelectric polarization.\cite{Fina} We, therefore, determined the intrinsic switchable ferroelectric polrization for this sample by measuring the remanent hysteresis loop. The remanent hysteresis loop is measured by eliminating the contribution of the nonremanent polarization by using the following protocol. First a preset voltage pulse comprising of only one half of the triangular wave (positive or negative) was used for prepolarizing the sample. Next a measurement pulse comprising of both positive and negative cycles was applied while data were measured only during one half of the pulse. The polarity of the measurement pulse was opposite to that of the preset pulse in order to switch the polarization of the sample. Then another train of preset and measurement pulses was used in which again the polarity of the measurement pulse was opposite to that of the preset pulse. For this second measurement pulse too, the data were recorded during only one half of the complete cycle. Combining the data obtained during the first and second measurement pulses (logic1), a complete hysteresis loop was obtained. This loop contains both the switchable and nonswitchable components of the polarization. Finally, two such preset and measurement pulses (logic0) were used to measure a complete hysteresis loop for only the nonswitchable component of polarization. In this case, the polarity of the preset and measurement pulses was kept the same. Subtraction of this latter loop from the former one yields the remanent hysteresis loop. This remanent hysteresis loop then measures the intrinsic hysteretic ferroelectric polarization of the sample. Comparison of the remanent hysteresis loop with the ordinary one is given in the supplementary document.\cite{supplementary}  The intrinsic remanent polarization turns out to be smaller. The train of voltage pulses used for measuring the remanent hysteresis loop is also given in the supplementary document.\cite{supplementary} The loops obtained from logic1 and logic0 pulse trains together with the remanent hysteresis loops are shown in the supplementary document\cite{supplementary} for a few representative cases. The measurement frequency also plays an important role as at higher frequency the influence of resistive leakage and interface effect minimizes.\cite{Fina} At a lower frequency, the remanent hysteresis loop exhibits an anomalous decrease in polarization with the increase in electric field. This could be due to an error in eliminating the influence of the nonswitchable polarization. At higher frequency, the error could be removed. In the supplementary document,\cite{supplementary} remanent hysteresis loops measured at different frequencies are shown. The remanent hysteresis loops being reported here were, therefore, measured at a time scale of 1 ms. We further show\cite{supplementary} that for a nonferroelectric perovskite sample Pr$_{0.5}$Ca$_{0.5}$MnO$_3$ at room temperature, measurement of remanent hysteresis loop using the above protocol yields virtually zero remanent polarization. Finite remanent polarization does imply presence of intrinsic hysteretic ferroelectricity in the sample. In Fig. 5a, we show the remanent hysteresis loops measured under zero and $\sim$15 kOe fields. The loop shape clearly proves the presence of finite remanent ferroelectric polarization in orthorhombic LuFeO$_3$. It is important to point out here that the remanent polarization increases with the decreases in temperature around the transition temperature T$_N$ ($\sim$600 K) and then reaches saturation, as expected, at even lower temperature (Fig. 2a). This pattern of variation of remanent polarization with temperature across a wide temperature range 100-700 K also proves the presence of genuine ferroelectricity in the sample. From Fig. 5a, it appears that a significant suppression of the remanent polarization takes place under a magnetic field of $\sim$15 kOe. The field (H) dependence of the extent of suppression of the remanent polarization $\frac{\Delta P_r}{P_r(0)}$ [=$\frac{P_r(0)-P_r(H)}{P_r(0)}$] is shown in Fig. 5b. Even though the polarization is small, substantial change under a magnetic field implies strong magnetoelectric multiferroic coupling. This has been observed in the case of other type-II multiferroics as well, such as DyMnO$_3$ or TbMnO$_3$.\cite{Goto}

\begin{figure}[!ht]
  \begin{center}
    \includegraphics[scale=0.42]{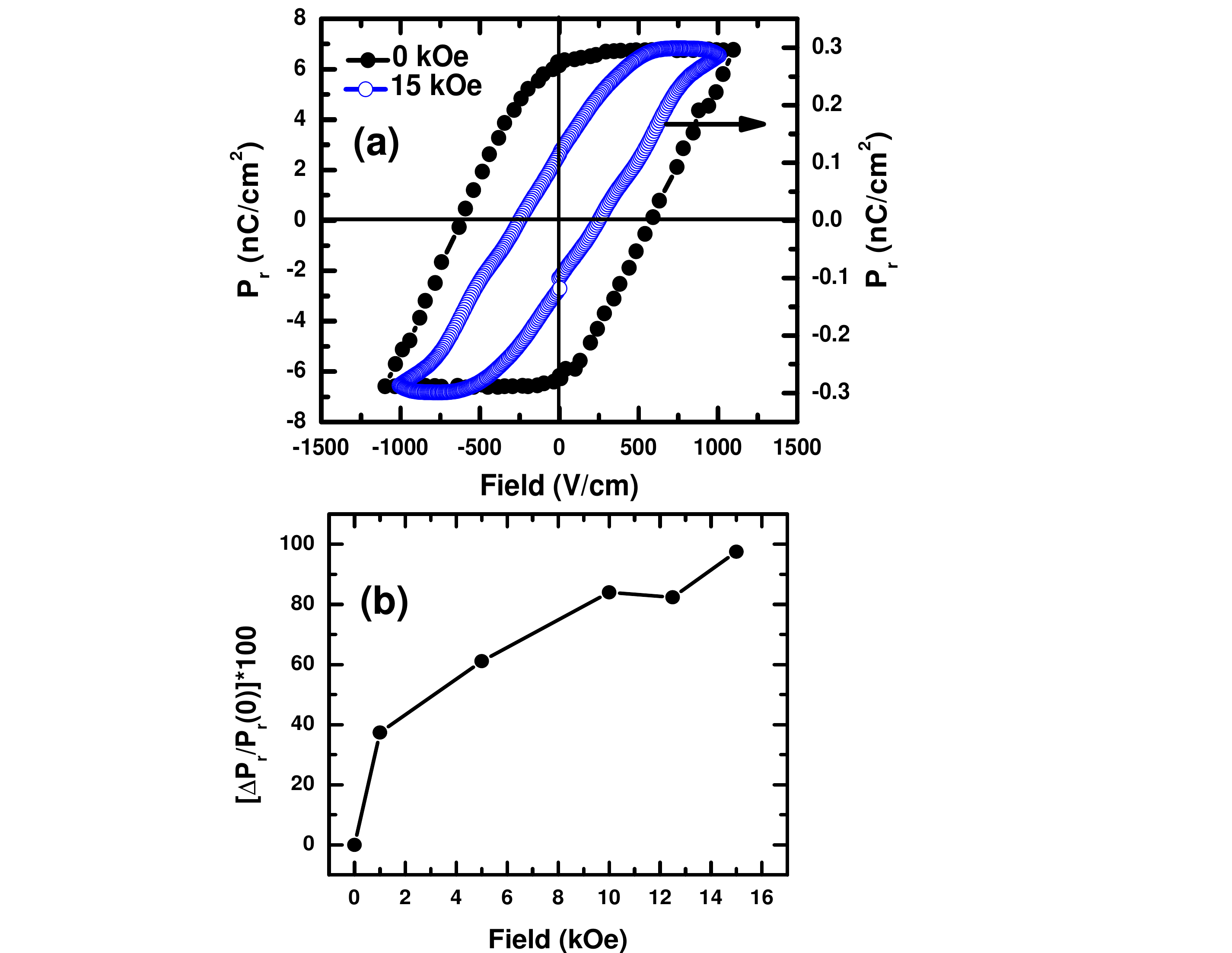} 
    \end{center}
\caption{(a) The remanent hysteresis loop under zero and $\sim$15 kOe magnetic field at room temperature; (b) the variation of the change in the remanent polarization with field.}
\end{figure}

From the theoretical calculation, the spin canting driven improper ferroelectricity, originating from the spin current\cite{Katsura} induced in noncollinear magnet via antisymmetric exchange coupling $\vec{\bf{S_1}} \times \vec{\bf{S_2}}$, was shown to yield a polarization of $\sim$10 nC/cm$^2$ at room temperature for isostructural SmFeO$_3$.\cite{Lee} The remanent polarization obtained here for LuFeO$_3$ (Fig. 5a), at room temperature, appears to be quite comparable to this theoretical prediction. The suppression of the remanent polarization under a magnetic field in orthorhombic LuFeO$_3$ could result from influence of magnetic field on the extent of noncollinearity of the magnetic structure which could be purely electronic or mixed electronic and ionic. For a well-known type-II multiferroic TbMnO$_3$, it has already been shown that magnetic field leads to displacement of the Tb, Mn, and O ions which, in turn, changes the polarization.\cite{Walker} Even though in a single crystal of TbMnO$_3$, a switching of polarization has been observed under the application of a critical magnetic field, substantial change in the polarization has been observed in the bulk polycrystalline sample as well.\cite{Staruch} This could result from incomplete switching of polarization under a given magnetic field (different from the critical field) which when summed over randomly oriented grains still yields a finite change. In the case of nanoscale BiFeO$_3$ too, from the neutron diffraction data under a magnetic field, it has been shown that ionic displacement indeed takes place under a magnetic field.\cite{Goswami} It is worth mentioning here that the magnetic structure in TbMnO$_3$, BiFeO$_3$, and LuFeO$_3$ are different. For LuFeO$_3$, significantly large change in electronic and ionic structure under a magnetic field could lead to the observed change in the remanent polarization. It has been shown here that the change in polarization is intrinsic as the influence of magnetic field on the resistive component is more than an order of magnitude weaker. Since in type-II multiferroics the magnetism and ferroelectricity are intimately coupled, strong magnetoelectric multiferroic coupling is expected. The results presented here show that the orthorhombic LuFeO$_3$ is another member of this type-II class. Hexagonal LuFeO$_3$, on the other hand, because of difference between the ferroelectric and magnetic transition points does not belong to the type-II variety however large the polarization might be. 

In summary, we report that we have observed finite ferroelectric polarization and strong multiferroic coupling at room temperature in orthorhombic LuFeO$_3$ from direct electrical measurements under a magnetic field. The polarization emerges here from the spin current, induced in noncollinear magnetic structure via antisymmetric exchange coupling interaction, and, therefore, both the ferroelectric and magnetic transitions take place simultaneously. The transitions, of course, are broad and taking place across a transition zone of $\sim$100 K. These results establish orthorhombic LuFeO$_3$ as a member of the type-II multiferroic family. Since the multiferroicity is observed at room temperature, LuFeO$_3$ could be extremely useful for spintronics based applications where spin structure is manipulated by electric field. 

\textbf{Acknowledgements.}
This work has been supported by the Department of Science and Technology(DST), Govt of India by a grant-in-aid project (SR/S2/CMP-0036/2009). One of the authors (UC) acknowledges support in the form of Senior Research Fellowship from DST. Another author (SG) acknowledges support in the form of a Research Associateship from CSIR, Govt of India.

\end{document}